\documentclass[twocolumn,showpacs,superscriptaddress,aps,prb]{revtex4-1}

\usepackage{graphicx}
\usepackage{bm}
\usepackage{epsfig}
\usepackage{epstopdf}
\usepackage{amsmath}

\begin{document}

\bibliographystyle{apsrev}

\title{Unusual size effects on thermoelectricity in a strongly correlated oxide}

\author{J. Ravichandran}
\email{jayakanth@berkeley.edu}
\affiliation{Applied Science and Technology Graduate Group, University of California, Berkeley, CA 94720, USA}
\affiliation{Materials Sciences Division, Lawrence Berkeley National Laboratory, Berkeley, CA 94720, USA}
\author{A. K. Yadav}
\affiliation{Materials Sciences Division, Lawrence Berkeley National Laboratory, Berkeley, CA 94720, USA}
\affiliation{Department of Materials Science and Engineering, University of California, Berkeley, CA 94720, USA}
\author{W. Siemons}
\affiliation{Materials Science and Technology Division, Oak Ridge National Laboratory, Oak Ridge, TN 37831, USA}
\author{M. A. McGuire}
\affiliation{Materials Science and Technology Division, Oak Ridge National Laboratory, Oak Ridge, TN 37831, USA}
\author{V. Wu}
\affiliation{Department of Materials Science and Engineering, University of California, Berkeley, CA 94720, USA}
\author{A. Vailionis}
\affiliation{Geballe Laboratory for Advanced Materials, Stanford University, Stanford, California 94305, USA}
\author{A. Majumdar}
\affiliation{ARPA-E, US Department of Energy, 1000 Independence Avenue, Washington, DC 20585, USA}
\author{R. Ramesh}	
\affiliation{Materials Sciences Division, Lawrence Berkeley National Laboratory, Berkeley, CA 94720, USA} 
\affiliation{Department of Materials Science and Engineering, University of California, Berkeley, CA 94720, USA}
\affiliation{Department of Physics, University of California, Berkeley, CA 94720, USA}
\affiliation{SETP, US Department of Energy, 1000 Independence Avenue, Washington, DC 20585, USA}

\date{\today}

\begin{abstract}
We investigated size effects on thermoelectricity in thin films of a strongly correlated layered cobaltate. At room temperature, the thermopower is independent of thickness down to 6 nm. This unusual behavior is inconsistent with the Fuchs-Sondheimer theory, which is used to describe conventional metals and semiconductors, and is attributed to the strong electron correlations in this material. Although the resistivity increases, as expected, below a critical thickness of $\sim$ 30 nm. The temperature dependent thermopower is similar for different thicknesses but resistivity shows systematic changes with thickness. Our experiments highlight the differences in thermoelectric behavior of strongly correlated and uncorrelated systems when subjected to finite size effects. We use the atomic limit Hubbard model at the high temperature limit to explain our observations. These findings provide new insights on decoupling electrical conductivity and thermopower in correlated systems. 
\end{abstract}


\maketitle 

\section{\label{sec:level1}Introduction}

Among various energy conversion methods, thermoelectricity deals with direct inter-conversion of thermal and electrical energy. The efficiency of a thermoelectric heat engine is related to a material dependent figure of merit, $Z$, given by $S^{2}\sigma/\kappa$, where $S$ is the thermopower or Seebeck coefficient, $\sigma$ and $\kappa$ are the electrical and thermal conductivities (lattice and electronic), respectively. In conventional thermoelectric materials, electrical conductivity and thermopower are governed by the density of states, chemical potential and the scattering mechanism. Due to this coupling between thermopower and electrical conductivity, achieving high $Z$ has been a challenging task. Hicks and Dresselhaus\cite{Hicks:1993tj,Hicks:1993tq} proposed quantum confinement as a means to enhance the thermoelectric power factor ($S^{2}\sigma$) in nanostructured materials. Nanostructuring\cite{Kim:2006tt,Poudel:2008cj} showed no significant enhancement in power factor, as the enhancement in thermopower was offset by the decrease in electrical conductivity (both mobility and carrier density). Nevertheless, nanostructuring is an effective means to reduce the lattice part of the thermal conductivity without significantly affecting electrical transport. Investigations exploring quantum confinement effects have primarily centered around conventional semiconductors, which show band-like transport.  

 Reduced dimensions in materials can have profound influence on transport properties due to effects such as quantization and changes in scattering mechanism. Thin films are the commonly used to study two dimensional transport behavior. There are several reports on thickness dependent transport measurements on thin film materials showing band-like transport,\cite{Pichard:1980vk,Worden:1958uj,Cho:2005tm,Ganesan:2000fl,Rogacheva:2003ut,Rogacheva:2002ej,Ohtomo:2004vj} but few studies focus on size effects on strongly correlated materials.\cite{Schultz:2009ib} Several of these investigations have centered around the effect of transverse confinement on thermoelectric transport. Recent investigations have shown large thermoelectric responses in complex oxides.\cite{Okuda:2001kr,Terasaki:1997kw,Lee:2006kg} Particularly, strongly correlated oxides such as cobaltates show enhanced thermopower, which cannot be explained by band-like transport. The transport behavior in these cobaltates has been explained by the Hubbard model, with the incorporation of spin degeneracy.\cite{Koshibae:2001jp,Wang:2003wx} Size dependent thermoelectric measurements on a strongly correlated cobaltate poses several interesting questions about the role of quantum confinement in thermoelectric transport, the effect of thickness on the mobility and spin degeneracy.  In this article, we report unusual size effects on thermoelectricity in a strongly correlated thermoelectric oxide, Bi$_2$Sr$_2$Co$_2$O$_y$ (abbreviated as BSCO) and use the Hubbard model to explain the physics behind the observations. We performed thermoelectric transport measurements in thin films of BSCO both as a function of thickness and temperature to elucidate size effects on this system and discuss possible directions for correlated thermoelectrics. 
 
 \section{\label{sec:level1}Experimental Methods}
  
 Thin films of BSCO (3 -- 170 nm thick) were grown on yttria stablized zirconia (YSZ) substrates using pulsed laser deposition from a stoichiometric ceramic target of BSCO. The growth was carried out at 700$^\circ$C, with an oxygen partial pressure of 500 mTorr. All films were grown using a 248 nm KrF excimer laser with a fluence of 2 J/cm$^2$ at a repetition rate of 1 Hz. Films were characterized by X-ray diffraction (XRD) for phase purity and crystallinity, X-ray reflectivity (XRR) for thickness, atomic force microscopy (AFM) for surface roughness and Rutherford backscattering (RBS) for chemical composition. All transport measurements were carried out in the van der Pauw geometry. Triangular ohmic metal contacts of side 1 mm (15 nm Ti/100 nm Pd) were deposited on the corners of the films using electron beam evaporation. Hall measurements at room temperature were carried out in air using a home-built apparatus with a 1 Tesla electromagnet. Low temperature resistivity and thermopower measurements were performed in a Quantum Design PPMS. 

\setlength{\epsfxsize}{1\columnwidth}
\begin{figure}[t]
\epsfbox{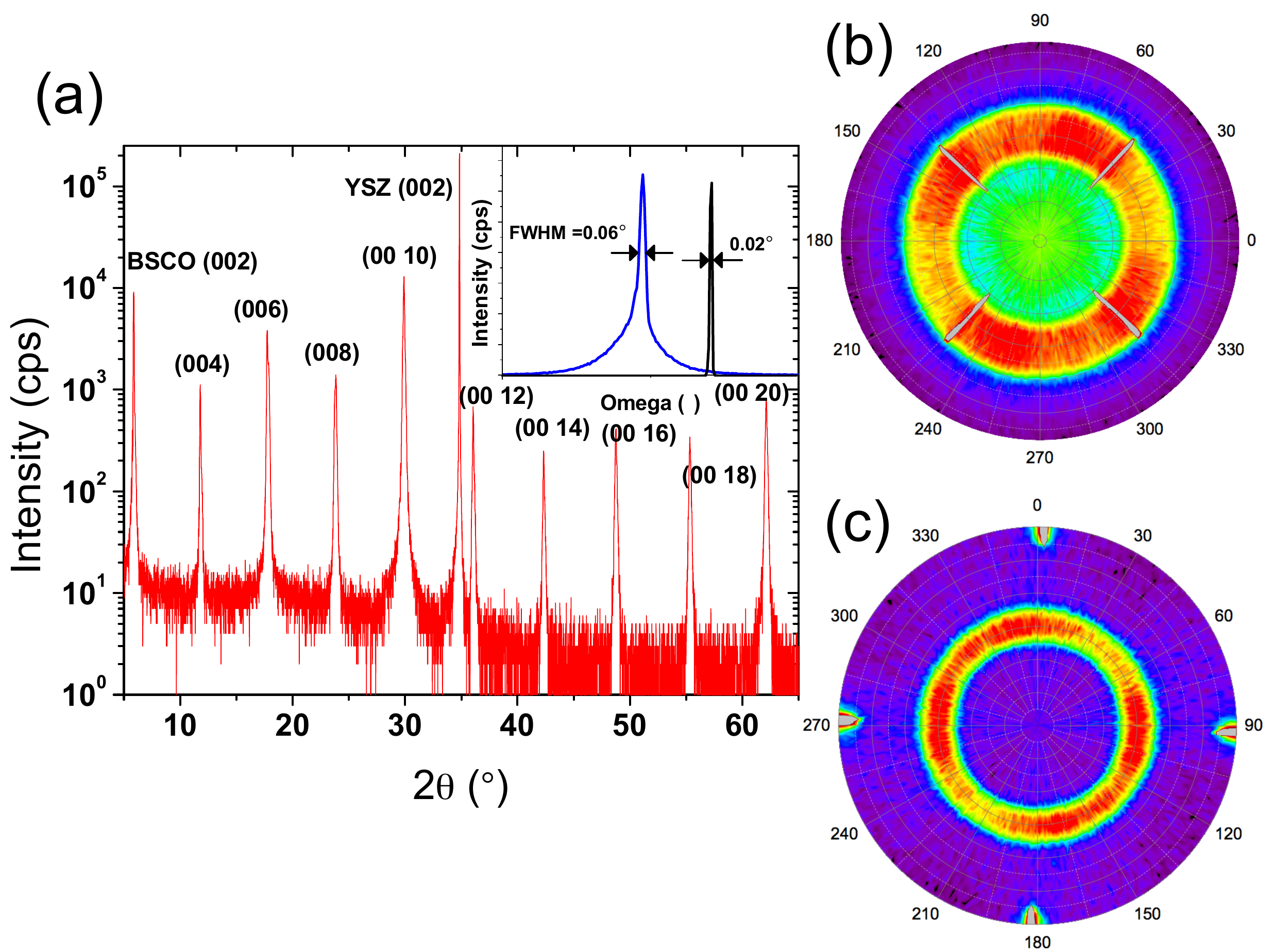}
\caption{\label{Fig:fig5} (a) The out-of-plane x-ray diffraction pattern for BSCO film on YSZ is shown. The inset compares the rocking curve for the substrate and the film. (b) Pole figure scan for (116) plane of BSCO. The four peaks at $\phi$ = 45$^\circ$, 135$^\circ$, 225$^\circ$ and 315$^\circ$ correspond to the substrate YSZ (111). The corresponding 2$\theta$ value for the scan is 29.75$^\circ$. (c) Pole figure scan for (11 14) plane of BSCO. The four peaks at $\phi$ = 0$^\circ$, 90$^\circ$, 180$^\circ$ and 270$^\circ$ correspond to the substrate YSZ (2 2 0). The corresponding 2$\theta$ value for the scan is 50.46$^\circ$.}
\end{figure}

\setlength{\epsfxsize}{1\columnwidth}
\begin{figure}[t]
\epsfbox{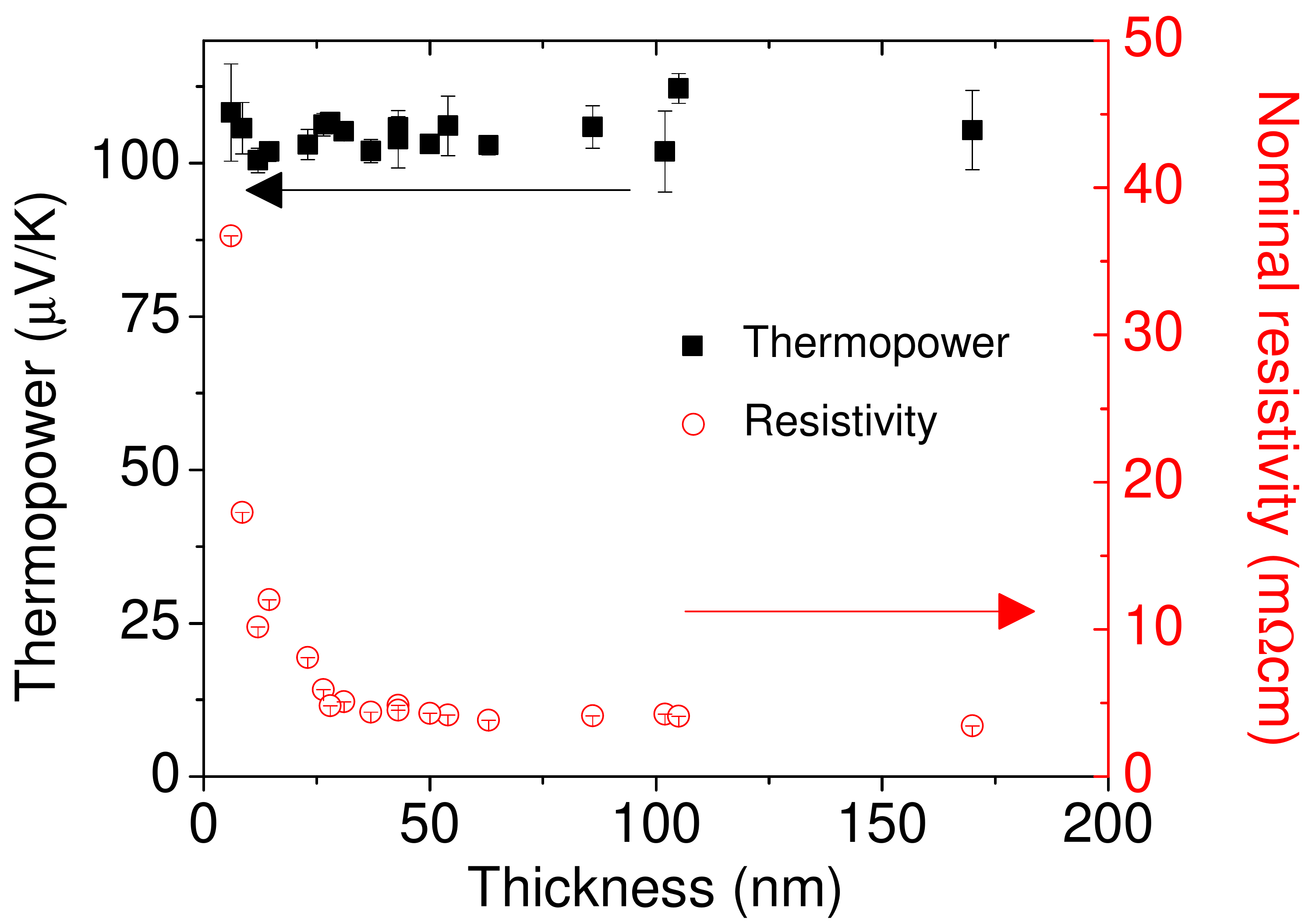}
\caption{\label{Fig:fig1} The thickness dependent thermopower and resistivity of thin films of BSCO measured at room temperature. The thicknesses of the films range from 6 -- 170 nm. The resistivity of the films was calculated by dividing the measured sheet resistance by the measured thickness. The actual resistivity can be estimated only after considering the thickness of dead layer present in these films, which we show in Fig.~\ref{Fig:fig2}, but the overall trend remains the same.}
\end{figure}

\section{\label{sec:level1}Results and Discussion}

The XRD measurements indicated that the films were single phase and oriented with c-axis (axis perpendicular to the layers) along the out-of-plane direction. The out-of-plane x-ray diffraction pattern for the BSCO film grown on YSZ is shown in Fig.~\ref{Fig:fig5}(a). All the peaks can be indexed to the (00$l$) planes of BSCO and no secondary phase or other orientations were observed. The inset of Fig.~\ref{Fig:fig5}(a) shows the rocking curve for the film as compared to the substrate. The films showed rocking curves with full-width-at-half-maximum of $\sim$ 0.05--0.2$^\circ$ (compared to substrate's 0.02$^\circ$). In order to establish the in-plane epitaxial relationship between the film and substrate, we performed phi-pole scans about (116) and (11 14) peaks of BSCO. The pole figures for the two cases are shown in Fig.~\ref{Fig:fig5}(b) and Fig.~\ref{Fig:fig5}(c) respectively. It is evident from the figures that the BSCO film doesnÕt have any preferential in-plane epitaxial relationship with the substrate. Thus, we have excellent out-of-plane texturing but no in-plane relationship with the substrate, leading to a Òwire textureÓ scenario. In Fig.~\ref{Fig:fig5}(b), four peaks at $\phi$ = 45$^\circ$, 135$^\circ$, 225$^\circ$ and 315$^\circ$ correspond to the substrate YSZ (111). The corresponding 2$\theta$ value for the scan is 29.75$^\circ$. In Fig.~\ref{Fig:fig5}(c), four peaks at $\phi$ = 0$^\circ$, 90$^\circ$, 180$^\circ$ and 270$^\circ$ correspond to the substrate YSZ (220). The corresponding 2$\theta$ value for the scan is 50.46$^\circ$.

 The primary result of this work is summarized in Fig.~\ref{Fig:fig1} where thickness dependent thermopower and resistivity for BSCO films at room temperature are shown. The nominal resistivity remained constant till $\sim$30 nm and for lower thicknesses the resistivity increased with decreasing thickness. Surprisingly, the thermopower remained constant ($\sim$100--110 $\mu$V/K) over the studied thickness range. In comparison, thermopower of  Se doped Bi$_2$Te$_3$ decreases by 65\% from the bulk value when the thickness is decreased to 50 nm.\cite{Cho:2005tm} Unlike BSCO, in a typical metal\cite{Pichard:1980vk,Worden:1958uj} or a semiconductor\cite{Cho:2005tm,Ganesan:2000fl}, decreasing thickness results in decrease in thermopower and increase in resistivity, due to surface scattering. In those systems, the observed thickness dependent thermoelectric properties can be explained by Fuchs-Sondheimer theory\cite{Sondheimer:1952ie} quantitatively for metals\cite{Pichard:1980vk,Worden:1958uj} and qualitatively for semiconductors.\cite{Cho:2005tm,Ganesan:2000fl,Rogacheva:2003ut} Fuchs-Sondheimer theory uses the energy dependent surface scattering as an additional scattering mechanism which becomes dominant when the thickness of the films are comparable to the bulk mean free path of the electrons. The thickness dependent thermopower and resistivity as predicted by this theory are given below.
 \setlength{\epsfxsize}{1\columnwidth}
\begin{figure}[t]
\epsfbox{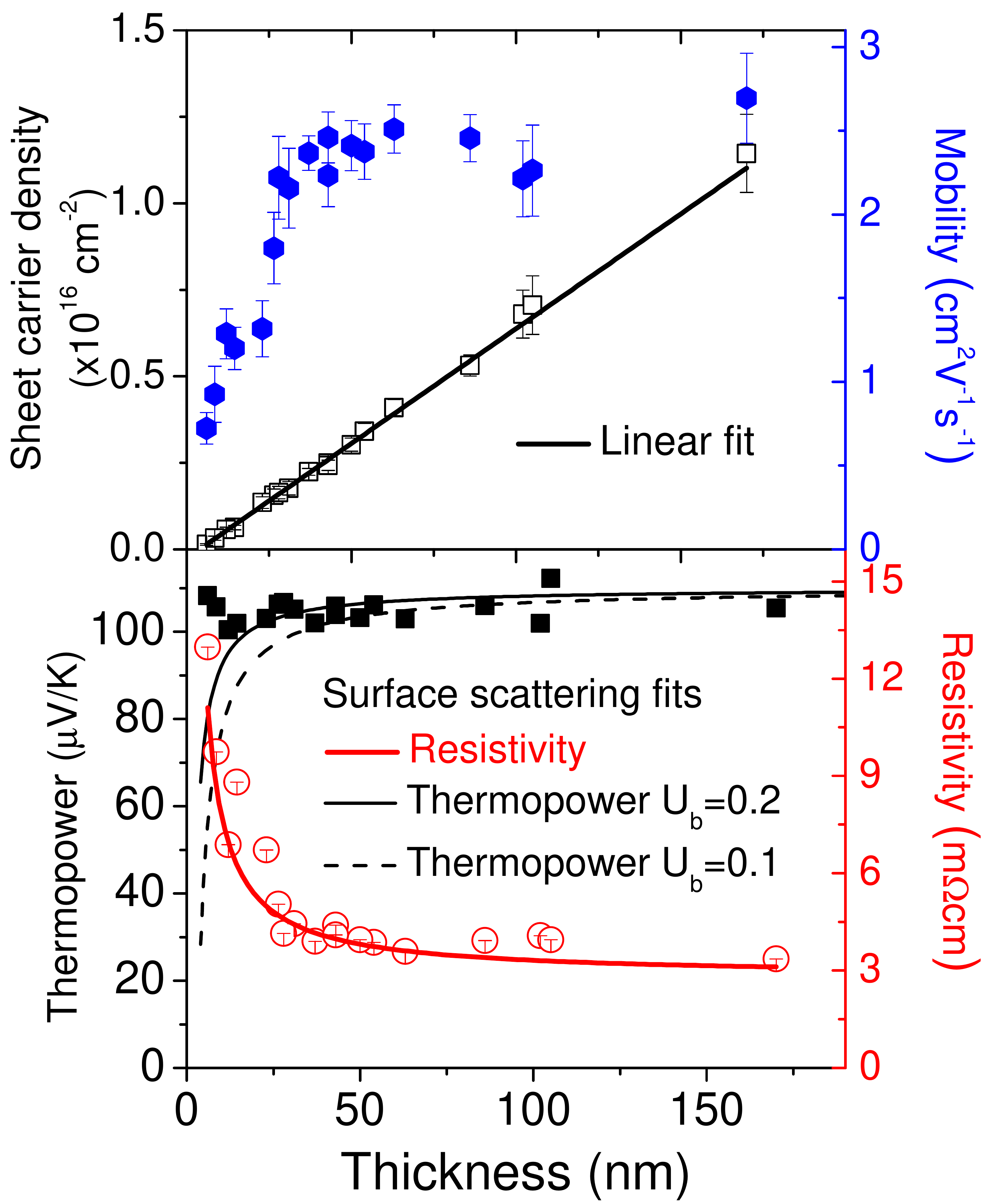}
\caption{\label{Fig:fig2} The sheet carrier density and Hall mobility as a function of thickness at room temperature. (Top panel) The sheet carrier density data was an excellent fit for a straight line with a small offset in thickness $\sim$ 4 nm. (Bottom panel) The thermopower  and the actual resistivity as a function of thickness, with the surface scatterings fits clearly showing the deviation for the thermopower data. The surface scattering model used the mean free path estimated from the resistivity data and typical values for the energy scattering dependent scattering term as 0.1 and 0.2.\cite{Cho:2005tm} The actual resistivity of the films is calculated accounting for the dead layer.}
\end{figure} 
 \begin{equation}
\rho_f = \rho_b\left(1+ \frac{3l_b}{8t}(1-p)\right)
\label{Eq:resistivity}
\end{equation}
\begin{equation}
S_f = S_b\left(1- \frac{3l_b}{8t}(1-p)\frac{U_b}{1+U_b}\right)
\label{Eq:thermopower}
\end{equation}
where $\rho_f$ is film resistivity, $\rho_b$ is bulk resistivity, $S_f$ is film thermopower, $S_b$ is bulk thermopower, $l_b$ is bulk electron mean free path, $t$ is the thickness, $p$ is the specularity parameter and $U_b$ is the energy dependent scattering term. If the Fuchs-Sondheimer model is applicable to our system, we expect a decrease in thermopower as thickness decreases, contradictory to the observed constant thermopower. We use in-depth transport measurements, to eliminate different scenarios under which this conventional theory can be applicable to our case and to establish the role of strong correlations in explaining our observations. 

\setlength{\epsfxsize}{1\columnwidth}
\begin{figure}[t]
\epsfbox{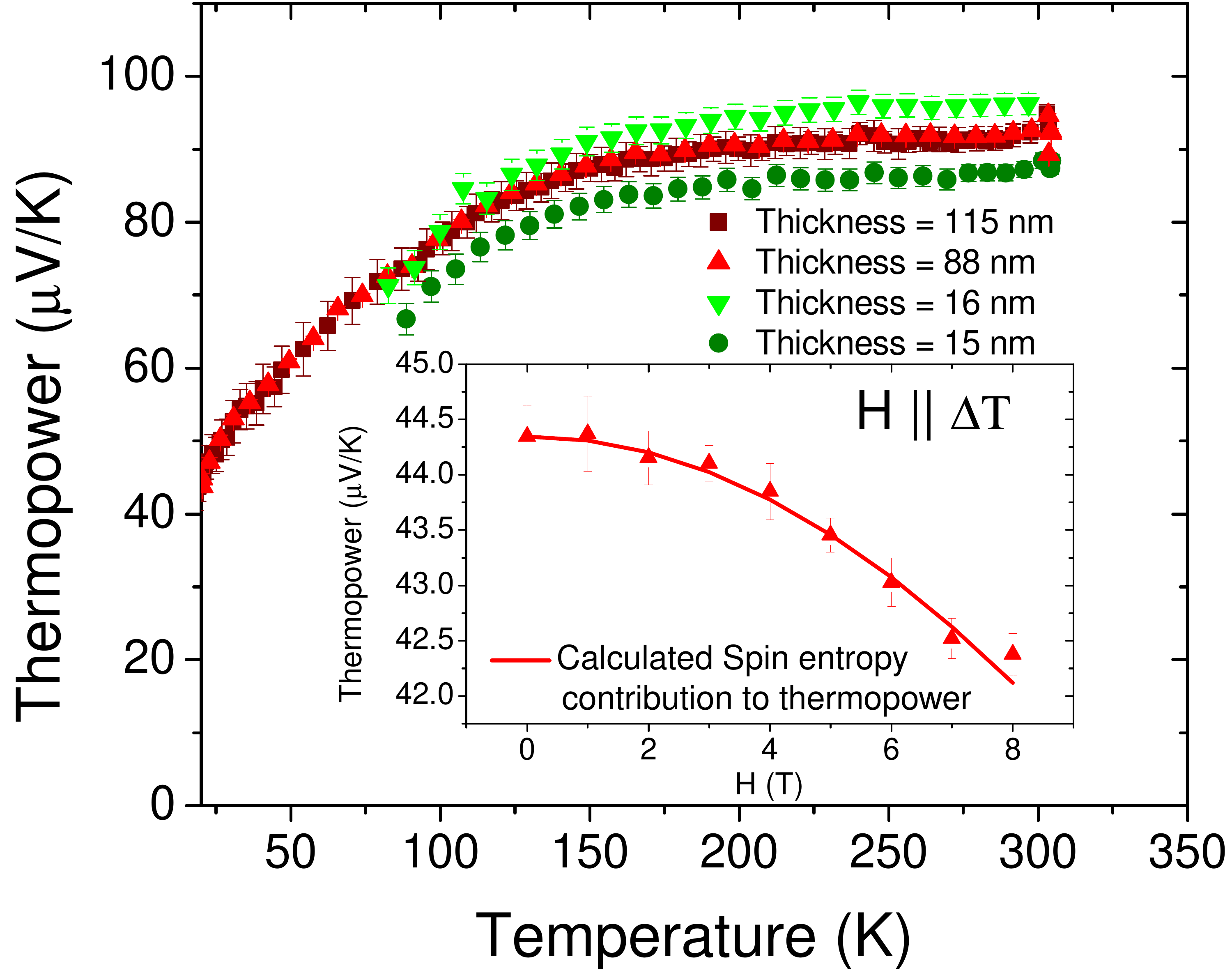}
\caption{\label{Fig:fig3} Temperature dependent thermopower for films of thickness 115, 88, 16 and 15 nm. The films showed very similar temperature dependence over the measured temperature range. The inset shows the magnetic field dependence of thermopower for a 88 nm film at 20 K with the field applied along the temperature gradient. The calculated spin entropic contribution to thermopower is shown as a red line.}
\end{figure}

 First, it is essential to understand the contribution of carrier concentration and mobility in increasing resistivity with decreasing thickness. Hall measurements were used to measure the sheet carrier density and mobility (see Fig.~\ref{Fig:fig2}). The sheet carrier density scaled linearly with thickness and has a small offset of $\sim$ 4 nm in the thickness axis. Thus, there is no thickness dependent change in the volume carrier density but an insulating dead layer of thickness $\sim$ 4 nm is present. We confirmed this by growing a film of $\sim$ 3 nm thickness and found it to be insulating. The measured Hall mobility showed a sharp decrease below 30 nm, as shown in Fig.~\ref{Fig:fig2}. Hence, we can conclude that the increase in resistivity shown in Fig.~\ref{Fig:fig1} is caused by the presence of a dead layer and the mobility reduction caused by surface scattering below $\sim$ 30 nm. We account for the presence of dead layer and plot the revised resistivity in Fig.~\ref{Fig:fig2}, which still shows very similar thickness dependence as depicted in Fig.~\ref{Fig:fig1} and an excellent fit for the surface scattering dominated resistivity shown in Eqn.\ref{Eq:resistivity}. On the other hand, the calculated thermopower values (using Eqn.\ref{Eq:thermopower}) with the mean free path values obtained from the resistivity fit and a typical energy dependence scattering term of 0.1 and 0.2, show clear deviations from the experimentally measured values below 50 and 30 nm respectively. 

 The presence of a competing mechanism such as quantum confinement,\cite{Hicks:1993tj}which compensates for the thermopower decrease due to surface scattering, can validate the applicability of Fuchs-Sondheimer theory. The lack of changes in the bulk carrier concentration is inconsistent with the presence of quantum confinement but is not a definitive proof to rule out this scenario completely. Typically, the surface scattering is not expected to show a strong temperature dependence as it is a boundary dominated mechanism. On the other hand, mechanisms like quantum confinement are expected to show a strong temperature dependence, as the effect is stronger at lower temperatures. Thus, if the temperature dependent thermopower does not show any significant deviation from the bulk at the lower thicknesses, we can conclude that Fuchs-Sondheimer theory is not applicable to our case and thermopower is insensitive to surface scattering. The temperature dependent thermopower as shown in Fig.~\ref{Fig:fig3} clearly depicts a very similar temperature dependence for thicknesses of 115, 88, 16 and 15 nm films down to 80 K (for thinner films the thermopower measurement became unreliable below this temperature due to high resistance of the films). Thus, the temperature dependence of thermopower remains bulk-like even in thin samples confirming that thermopower is robust against surface scattering and the non-applicability of Fuchs-Sondheimer theory in this system. It is important to note that the presence of wire-texture in these films, suggests that grain boundary scattering should also be considered for the electron scattering mechanisms. Typically, grain boundary scattering doesn't show any thickness dependence,\cite{Pichard:1980vk,Worden:1958uj} hence, will not change the conclusions derived here.
 
 The magnetic field dependence of thermopower for an 88 nm film at 20 K is shown in the inset of Fig.~\ref{Fig:fig3}. The observed field dependent thermopower is in excellent agreement with the spin entropic contribution to thermopower (Eqn.~\ref{Eq:Spin}).
\begin{equation}
Q(H,T)/Q(0,T) = {ln[2cosh(u)] - utanh(u)}/ln(2)
\label{Eq:Spin}
\end{equation}
where u = $g\mu_{B}$H/2k$_{\mathrm{B}}T$ and $g$ is the Land\'e $g$-factor (here, $g$=2). This observation is consistent with the experiments on Na$_x$CoO$_2$ ($g$=2.2),\cite{Wang:2003wx} confirming the role of strong correlation and spin entropy in thermoelectric properties of BSCO.
\setlength{\epsfxsize}{1\columnwidth}
\begin{figure}[t]
\epsfbox{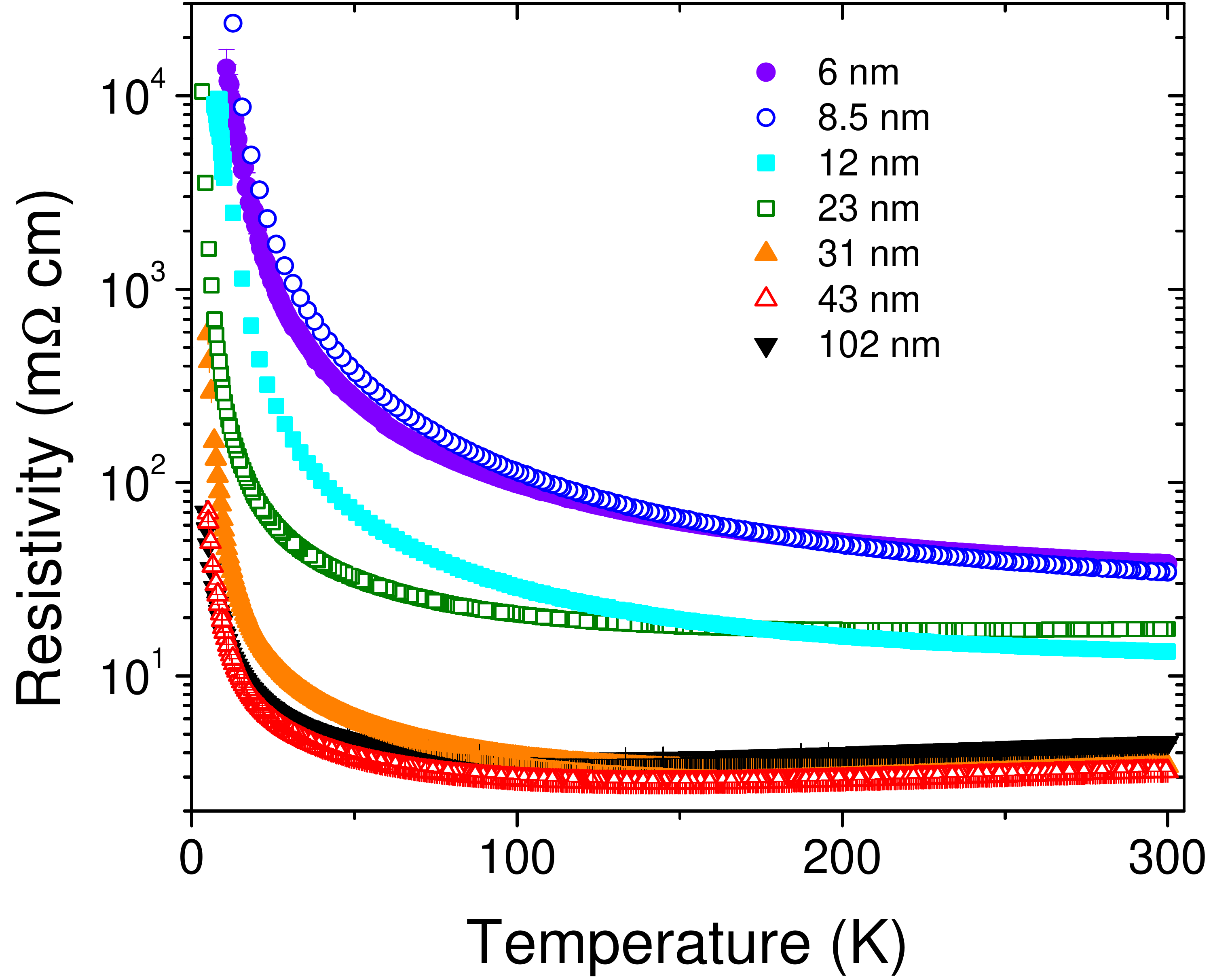}
\caption{\label{Fig:fig4} Temperature dependent resistivity for films of different thickness. At the bulk limit, the films show the characteristic metal--insulator transition with a transition temperature $\sim$ 100 K. As the thickness is decreased the transition temperature shifts to higher temperature and below 23 nm, the films remained insulating till 300 K.}
\end{figure}
 
  Finally, we studied the size effects on resistivity at low temperatures by performing temperature dependent resistivity measurements (shown in Fig.~\ref{Fig:fig4}) on films with different thicknesses. Single crystals\cite{Fujii:2002vd,Yamamoto:2002ic} and thin films\cite{Wang:2009eaa,Wang:2009hj} of BSCO have shown a metal-insulator transition with transition temperatures $\sim$80--140 K. Interestingly, the transition temperature shifted to higher temperatures as we decreased the thickness. This shift in transition temperature needs further investigation for a clear understanding. Moreover, due to the presence of unconventional Hall effect\cite{Eng:2006vc} in the cobaltates at low temperatures, extensive Hall effect investigations are necessary to uncover the exact origin of this shift.   
 
 As we have already established that BSCO is also a correlated system similar to other cobaltates, it is essential to put these findings in perspective within the framework of the Hubbard model.\cite{Koshibae:2000vj,Uchida:2011fv,Koshibae:2001jp,Mukerjee:2007tg} The transport coefficients predicted for Na$_x$CoO$_2$\cite{Mukerjee:2007tg} using the atomic limit Hubbard model in the high temperature limit are given as:

\begin{equation}
S = -\frac{k_B}{e} log \left(\frac{2(1-x)}{x}\right)
\label{Eq:Hthermopower}
\end{equation}
\begin{equation}
\rho = \frac{Vh^2}{8\pi^2e^{2}\eta a^2 t^2 \tau\beta x(1-x)}
\label{Eq:Hresistivity}
\end{equation}
where $x$ is filling, k$_B$ is the Boltzmann constant, e is the electronic charge, $\beta$ is 1/k$_{\mathrm{B}}T$, $\eta$ is the lattice structure dependent constant, $a$ is the lattice constant, $t$ is the bandwidth, $\tau$ is the relaxation time, $V$ is the unit cell volume and h is Planck's constant. 

The thermopower relation shown in Eqn.~\ref{Eq:Hthermopower} does not depend on the relaxation time and hence is consistent with our conclusion that thermopower is robust to changes in the scattering mechanism. Besides, the resistivity clearly shows an inverse scaling with the scattering time, hence consistent with our observations. Thus, it is evident that the Fuchs-Sondheimer theory doesn't account for the thermopower and resistivity measurements, but the simple Hubbard based clearly explains the thickness and temperature dependence of thermoelectric properties in this strongly correlated system. It is important to note that Eqn.\ref{Eq:Hthermopower} and Eqn.~\ref{Eq:Hresistivity} change qualitatively, if the limits on the energy scales such as thermal energy($k_{B}T$) and bandwidth ($t$) are different from the assumed limit here ($t \ll k_{B}T$). These variations still doesn't change the overall conclusion that thermopower is independent of scattering time. Our experiment elegantly establishes that the thermopower, at the high temperature limit, is independent of scattering parameter. It is important to comment on the relevance of the Eqn.\ref{Eq:Hthermopower}, in estimating the valence state of Co. 	Using the average room temperature thermopower of 110 $\mu V/K$, we estimate the x to be 0.36. Thus the estimated average cobalt valence in this compound is 3.36, which is very close to the reported value of 3.3.\cite{Morita:2004ty} Further, the resistivity in the metallic regime can be explained using Eqn.~\ref{Eq:Hresistivity} but there is no clear insight on how the metal insulator transition can be understood using the Hubbard model. 

\section{\label{sec:level1}Summary}
 
In summary, we have studied the size effects on thermoelectricity in thin films of a strongly correlated cobaltate system. The thermopower is insensitive to surface scattering unlike resistivity, which increases with decreasing thickness below $\sim$ 30 nm. These observations can be explained by the atomic limit Hubbard model. Unlike conventional thermoelectric materials, the insensitivity of thermopower to scattering mechanism in strongly correlated systems simplifies the decoupling of thermopower and electrical conductivity. Hence, the next step towards complete decoupling of thermopower and electrical conductivity in a correlated system is only dependent on understanding the limits of filling dependence of thermoelectric properties. Since nanostructuring is a proven route to decrease the lattice part of thermal conductivity without affecting electrical properties, designing nanostructured correlated materials can lead to decoupling of all the three thermoelectric parameters and hence, a pathway to high thermoelectric efficiency.

\begin{acknowledgments}
The work was supported by the Division of Materials Sciences and Engineering, Office of Basic Energy Sciences, U.S Department of Energy. JR acknowledges support from the Link foundation. The authors gratefully acknowledge the assistance of Dr. K. M. Yu with the RBS measurements, target preparation by J. Wu and useful discussions with Dr. S. Mukerjee, Dr. Jay Sau, Dr. B. Kavaipatti, Dr. M. Trassin, Dr. C.-W Liang and Dr. Y.-H Chu. 
\end{acknowledgments}


\begin{thebibliography}{26}
\expandafter\ifx\csname natexlab\endcsname\relax\def\natexlab#1{#1}\fi
\expandafter\ifx\csname bibnamefont\endcsname\relax
  \def\bibnamefont#1{#1}\fi
\expandafter\ifx\csname bibfnamefont\endcsname\relax
  \def\bibfnamefont#1{#1}\fi
\expandafter\ifx\csname citenamefont\endcsname\relax
  \def\citenamefont#1{#1}\fi
\expandafter\ifx\csname url\endcsname\relax
  \def\url#1{\texttt{#1}}\fi
\expandafter\ifx\csname urlprefix\endcsname\relax\def\urlprefix{URL }\fi
\providecommand{\bibinfo}[2]{#2}
\providecommand{\eprint}[2][]{\url{#2}}

\bibitem[{\citenamefont{Hicks and
  Dresselhaus}(1993{\natexlab{a}})}]{Hicks:1993tj}
\bibinfo{author}{\bibfnamefont{L.}~\bibnamefont{Hicks}} \bibnamefont{and}
  \bibinfo{author}{\bibfnamefont{M.}~\bibnamefont{Dresselhaus}},
  \bibinfo{journal}{Physical Review B} \textbf{\bibinfo{volume}{47}},
  \bibinfo{pages}{12727} (\bibinfo{year}{1993}{\natexlab{a}}).

\bibitem[{\citenamefont{Hicks and
  Dresselhaus}(1993{\natexlab{b}})}]{Hicks:1993tq}
\bibinfo{author}{\bibfnamefont{L.}~\bibnamefont{Hicks}} \bibnamefont{and}
  \bibinfo{author}{\bibfnamefont{M.}~\bibnamefont{Dresselhaus}},
  \bibinfo{journal}{Physical Review B} \textbf{\bibinfo{volume}{47}},
  \bibinfo{pages}{16631} (\bibinfo{year}{1993}{\natexlab{b}}).

\bibitem[{\citenamefont{Kim et~al.}(2006)\citenamefont{Kim, Zide, Gossard,
  Klenov, and Stemmer}}]{Kim:2006tt}
\bibinfo{author}{\bibfnamefont{W.}~\bibnamefont{Kim}},
  \bibinfo{author}{\bibfnamefont{J.}~\bibnamefont{Zide}},
  \bibinfo{author}{\bibfnamefont{A.}~\bibnamefont{Gossard}},
  \bibinfo{author}{\bibfnamefont{D.}~\bibnamefont{Klenov}}, \bibnamefont{and}
  \bibinfo{author}{\bibfnamefont{S.}~\bibnamefont{Stemmer}},
  \bibinfo{journal}{Physical Review Letters} \textbf{\bibinfo{volume}{96}},
  \bibinfo{pages}{045901} (\bibinfo{year}{2006}).

\bibitem[{\citenamefont{Poudel et~al.}(2008)\citenamefont{Poudel, Hao, Ma, Lan,
  Minnich, Yu, Yan, Wang, Muto, Vashaee et~al.}}]{Poudel:2008cj}
\bibinfo{author}{\bibfnamefont{B.}~\bibnamefont{Poudel}},
  \bibinfo{author}{\bibfnamefont{Q.}~\bibnamefont{Hao}},
  \bibinfo{author}{\bibfnamefont{Y.}~\bibnamefont{Ma}},
  \bibinfo{author}{\bibfnamefont{Y.}~\bibnamefont{Lan}},
  \bibinfo{author}{\bibfnamefont{A.}~\bibnamefont{Minnich}},
  \bibinfo{author}{\bibfnamefont{B.}~\bibnamefont{Yu}},
  \bibinfo{author}{\bibfnamefont{X.}~\bibnamefont{Yan}},
  \bibinfo{author}{\bibfnamefont{D.}~\bibnamefont{Wang}},
  \bibinfo{author}{\bibfnamefont{A.}~\bibnamefont{Muto}},
  \bibinfo{author}{\bibfnamefont{D.}~\bibnamefont{Vashaee}},
  \bibnamefont{et~al.}, \bibinfo{journal}{Science}
  \textbf{\bibinfo{volume}{320}}, \bibinfo{pages}{634} (\bibinfo{year}{2008}).
  
  \bibitem[{\citenamefont{Pichard et~al.}(1980)\citenamefont{Pichard, Tellier,
  and Tosser}}]{Pichard:1980vk}
\bibinfo{author}{\bibfnamefont{C.}~\bibnamefont{Pichard}},
  \bibinfo{author}{\bibfnamefont{C.}~\bibnamefont{Tellier}}, \bibnamefont{and}
  \bibinfo{author}{\bibfnamefont{A.}~\bibnamefont{Tosser}},
  \bibinfo{journal}{Journal of Physics F: Metal Physics}
  \textbf{\bibinfo{volume}{10}}, \bibinfo{pages}{2009} (\bibinfo{year}{1980}).

\bibitem[{\citenamefont{Worden}(1958)}]{Worden:1958uj}
\bibinfo{author}{\bibfnamefont{D.}~\bibnamefont{Worden}},
  \bibinfo{journal}{Journal Of Physics And Chemistry Of Solids}
  \textbf{\bibinfo{volume}{6}}, \bibinfo{pages}{89} (\bibinfo{year}{1958}).

\bibitem[{\citenamefont{Cho and Kim}(2005)}]{Cho:2005tm}
\bibinfo{author}{\bibfnamefont{K.}~\bibnamefont{Cho}} \bibnamefont{and}
  \bibinfo{author}{\bibfnamefont{I.}~\bibnamefont{Kim}},
  \bibinfo{journal}{Materials Letters} \textbf{\bibinfo{volume}{59}},
  \bibinfo{pages}{966} (\bibinfo{year}{2005}).

\bibitem[{\citenamefont{Ganesan and Sivaramakrishnan}(2000)}]{Ganesan:2000fl}
\bibinfo{author}{\bibfnamefont{N.}~\bibnamefont{Ganesan}} \bibnamefont{and}
  \bibinfo{author}{\bibfnamefont{V.}~\bibnamefont{Sivaramakrishnan}},
  \bibinfo{journal}{Journal Of Physics D-Applied Physics}
  \textbf{\bibinfo{volume}{21}}, \bibinfo{pages}{784} (\bibinfo{year}{2000}).

\bibitem[{\citenamefont{Rogacheva et~al.}(2003)\citenamefont{Rogacheva,
  Nashchekina, Vekhov, Dresselhaus, and Cronin}}]{Rogacheva:2003ut}
\bibinfo{author}{\bibfnamefont{E.}~\bibnamefont{Rogacheva}},
  \bibinfo{author}{\bibfnamefont{O.}~\bibnamefont{Nashchekina}},
  \bibinfo{author}{\bibfnamefont{Y.}~\bibnamefont{Vekhov}},
  \bibinfo{author}{\bibfnamefont{M.}~\bibnamefont{Dresselhaus}},
  \bibnamefont{and} \bibinfo{author}{\bibfnamefont{S.}~\bibnamefont{Cronin}},
  \bibinfo{journal}{Thin Solid Films} \textbf{\bibinfo{volume}{423}},
  \bibinfo{pages}{115} (\bibinfo{year}{2003}).

\bibitem[{\citenamefont{Rogacheva et~al.}(2002)\citenamefont{Rogacheva,
  Tavrina, Grigorov, Nashchekina, Volobuev, Fedorov, Nasedkin, and
  Dresselhaus}}]{Rogacheva:2002ej}
\bibinfo{author}{\bibfnamefont{E.~I.} \bibnamefont{Rogacheva}},
  \bibinfo{author}{\bibfnamefont{T.~V.} \bibnamefont{Tavrina}},
  \bibinfo{author}{\bibfnamefont{S.~N.} \bibnamefont{Grigorov}},
  \bibinfo{author}{\bibfnamefont{O.~N.} \bibnamefont{Nashchekina}},
  \bibinfo{author}{\bibfnamefont{V.~V.} \bibnamefont{Volobuev}},
  \bibinfo{author}{\bibfnamefont{A.~G.} \bibnamefont{Fedorov}},
  \bibinfo{author}{\bibfnamefont{K.~A.} \bibnamefont{Nasedkin}},
  \bibnamefont{and} \bibinfo{author}{\bibfnamefont{M.~S.}
  \bibnamefont{Dresselhaus}}, \bibinfo{journal}{Journal of Electronic
  Materials} \textbf{\bibinfo{volume}{31}}, \bibinfo{pages}{298}
  (\bibinfo{year}{2002}).

\bibitem[{\citenamefont{Ohtomo and Hwang}(2004)}]{Ohtomo:2004vj}
\bibinfo{author}{\bibfnamefont{A.}~\bibnamefont{Ohtomo}} \bibnamefont{and}
  \bibinfo{author}{\bibfnamefont{H.}~\bibnamefont{Hwang}},
  \bibinfo{journal}{Applied Physics Letters} \textbf{\bibinfo{volume}{84}},
  \bibinfo{pages}{1716} (\bibinfo{year}{2004}).

\bibitem[{\citenamefont{Schultz et~al.}(2009)\citenamefont{Schultz, Levy,
  Reiner, and Klein}}]{Schultz:2009ib}
\bibinfo{author}{\bibfnamefont{M.}~\bibnamefont{Schultz}},
  \bibinfo{author}{\bibfnamefont{S.}~\bibnamefont{Levy}},
  \bibinfo{author}{\bibfnamefont{J.}~\bibnamefont{Reiner}}, \bibnamefont{and}
  \bibinfo{author}{\bibfnamefont{L.}~\bibnamefont{Klein}},
  \bibinfo{journal}{Physical Review B} \textbf{\bibinfo{volume}{79}},
  \bibinfo{pages}{125444} (\bibinfo{year}{2009}).

\bibitem[{\citenamefont{Okuda et~al.}(2001)\citenamefont{Okuda, Nakanishi,
  Miyasaka, and Tokura}}]{Okuda:2001kr}
\bibinfo{author}{\bibfnamefont{T.}~\bibnamefont{Okuda}},
  \bibinfo{author}{\bibfnamefont{K.}~\bibnamefont{Nakanishi}},
  \bibinfo{author}{\bibfnamefont{S.}~\bibnamefont{Miyasaka}}, \bibnamefont{and}
  \bibinfo{author}{\bibfnamefont{Y.}~\bibnamefont{Tokura}},
  \bibinfo{journal}{Physical Review B} \textbf{\bibinfo{volume}{63}},
  \bibinfo{pages}{113104} (\bibinfo{year}{2001}).

\bibitem[{\citenamefont{Wang et~al.}(2003)\citenamefont{Wang, Rogado, Cava, and
  Ong}}]{Wang:2003wx}
\bibinfo{author}{\bibfnamefont{Y.}~\bibnamefont{Wang}},
  \bibinfo{author}{\bibfnamefont{N.}~\bibnamefont{Rogado}},
  \bibinfo{author}{\bibfnamefont{R.}~\bibnamefont{Cava}}, \bibnamefont{and}
  \bibinfo{author}{\bibfnamefont{N.}~\bibnamefont{Ong}},
  \bibinfo{journal}{Nature} \textbf{\bibinfo{volume}{423}},
  \bibinfo{pages}{425} (\bibinfo{year}{2003}).

\bibitem[{\citenamefont{Terasaki et~al.}(1997)\citenamefont{Terasaki, Sasago,
  and Uchinokura}}]{Terasaki:1997kw}
\bibinfo{author}{\bibfnamefont{I.}~\bibnamefont{Terasaki}},
  \bibinfo{author}{\bibfnamefont{Y.}~\bibnamefont{Sasago}}, \bibnamefont{and}
  \bibinfo{author}{\bibfnamefont{K.}~\bibnamefont{Uchinokura}},
  \bibinfo{journal}{Physical Review B} \textbf{\bibinfo{volume}{56}},
  \bibinfo{pages}{R12685} (\bibinfo{year}{1997}).

\bibitem[{\citenamefont{Lee et~al.}(2006)\citenamefont{Lee, Viciu, Li, Wang,
  Foo, Watauchi, Pascal, Cava, and Ong}}]{Lee:2006kg}
\bibinfo{author}{\bibfnamefont{M.}~\bibnamefont{Lee}},
  \bibinfo{author}{\bibfnamefont{L.}~\bibnamefont{Viciu}},
  \bibinfo{author}{\bibfnamefont{L.}~\bibnamefont{Li}},
  \bibinfo{author}{\bibfnamefont{Y.}~\bibnamefont{Wang}},
  \bibinfo{author}{\bibfnamefont{M.}~\bibnamefont{Foo}},
  \bibinfo{author}{\bibfnamefont{S.}~\bibnamefont{Watauchi}},
  \bibinfo{author}{\bibfnamefont{R.}~\bibnamefont{Pascal}},
  \bibinfo{author}{\bibfnamefont{R.}~\bibnamefont{Cava}}, \bibnamefont{and}
  \bibinfo{author}{\bibfnamefont{N.}~\bibnamefont{Ong}},
  \bibinfo{journal}{Nature Materials} \textbf{\bibinfo{volume}{5}},
  \bibinfo{pages}{537} (\bibinfo{year}{2006}).

\bibitem[{\citenamefont{Koshibae and Maekawa}(2001)}]{Koshibae:2001jp}
\bibinfo{author}{\bibfnamefont{W.}~\bibnamefont{Koshibae}} \bibnamefont{and}
  \bibinfo{author}{\bibfnamefont{S.}~\bibnamefont{Maekawa}},
  \bibinfo{journal}{Physical Review Letters} \textbf{\bibinfo{volume}{87}},
  \bibinfo{pages}{236603} (\bibinfo{year}{2001}).

\bibitem[{\citenamefont{Sondheimer}(1952)}]{Sondheimer:1952ie}
\bibinfo{author}{\bibfnamefont{E.~H.} \bibnamefont{Sondheimer}},
  \bibinfo{journal}{Advances In Physics} \textbf{\bibinfo{volume}{1}},
  \bibinfo{pages}{1} (\bibinfo{year}{1952}).

\bibitem[{\citenamefont{Fujii et~al.}(2002)\citenamefont{Fujii, Terasaki,
  Watanabe, and Matsuda}}]{Fujii:2002vd}
\bibinfo{author}{\bibfnamefont{T.}~\bibnamefont{Fujii}},
  \bibinfo{author}{\bibfnamefont{I.}~\bibnamefont{Terasaki}},
  \bibinfo{author}{\bibfnamefont{T.}~\bibnamefont{Watanabe}}, \bibnamefont{and}
  \bibinfo{author}{\bibfnamefont{A.}~\bibnamefont{Matsuda}},
  \bibinfo{journal}{Japanese Journal Of Applied Physics Part 2-Letters {\&}
  Express Letters} \textbf{\bibinfo{volume}{41}}, \bibinfo{pages}{783}
  (\bibinfo{year}{2002}).

\bibitem[{\citenamefont{Yamamoto et~al.}(2002)\citenamefont{Yamamoto,
  Uchinokura, and Tsukada}}]{Yamamoto:2002ic}
\bibinfo{author}{\bibfnamefont{T.}~\bibnamefont{Yamamoto}},
  \bibinfo{author}{\bibfnamefont{K.}~\bibnamefont{Uchinokura}},
  \bibnamefont{and} \bibinfo{author}{\bibfnamefont{I.}~\bibnamefont{Tsukada}},
  \bibinfo{journal}{Physical Review B} \textbf{\bibinfo{volume}{65}},
  \bibinfo{pages}{184434} (\bibinfo{year}{2002}).

\bibitem[{\citenamefont{Wang et~al.}(2009{\natexlab{a}})\citenamefont{Wang,
  Venimadhav, Guo, Chen, Li, Soukiassian, Schlom, Katz, Pan, Wong-Ng
  et~al.}}]{Wang:2009eaa}
\bibinfo{author}{\bibfnamefont{S.}~\bibnamefont{Wang}},
  \bibinfo{author}{\bibfnamefont{A.}~\bibnamefont{Venimadhav}},
  \bibinfo{author}{\bibfnamefont{S.}~\bibnamefont{Guo}},
  \bibinfo{author}{\bibfnamefont{K.}~\bibnamefont{Chen}},
  \bibinfo{author}{\bibfnamefont{Q.}~\bibnamefont{Li}},
  \bibinfo{author}{\bibfnamefont{A.}~\bibnamefont{Soukiassian}},
  \bibinfo{author}{\bibfnamefont{D.~G.} \bibnamefont{Schlom}},
  \bibinfo{author}{\bibfnamefont{M.~B.} \bibnamefont{Katz}},
  \bibinfo{author}{\bibfnamefont{X.~Q.} \bibnamefont{Pan}},
  \bibinfo{author}{\bibfnamefont{W.}~\bibnamefont{Wong-Ng}},
  \bibnamefont{et~al.}, \bibinfo{journal}{Applied Physics Letters}
  \textbf{\bibinfo{volume}{94}}, \bibinfo{pages}{022110}
  (\bibinfo{year}{2009}{\natexlab{a}}).

\bibitem[{\citenamefont{Wang et~al.}(2009{\natexlab{b}})\citenamefont{Wang,
  Zhang, He, Chen, Yu, and Fu}}]{Wang:2009hj}
\bibinfo{author}{\bibfnamefont{S.}~\bibnamefont{Wang}},
  \bibinfo{author}{\bibfnamefont{Z.}~\bibnamefont{Zhang}},
  \bibinfo{author}{\bibfnamefont{L.}~\bibnamefont{He}},
  \bibinfo{author}{\bibfnamefont{M.}~\bibnamefont{Chen}},
  \bibinfo{author}{\bibfnamefont{W.}~\bibnamefont{Yu}}, \bibnamefont{and}
  \bibinfo{author}{\bibfnamefont{G.}~\bibnamefont{Fu}},
  \bibinfo{journal}{Applied Physics Letters} \textbf{\bibinfo{volume}{94}},
  \bibinfo{pages}{162108} (\bibinfo{year}{2009}{\natexlab{b}}).

\bibitem[{\citenamefont{Eng et~al.}(2006)\citenamefont{Eng, Limelette,
  Prellier, Simon, and Fr{\'e}sard}}]{Eng:2006vc}
\bibinfo{author}{\bibfnamefont{H.}~\bibnamefont{Eng}},
  \bibinfo{author}{\bibfnamefont{P.}~\bibnamefont{Limelette}},
  \bibinfo{author}{\bibfnamefont{W.}~\bibnamefont{Prellier}},
  \bibinfo{author}{\bibfnamefont{C.}~\bibnamefont{Simon}}, \bibnamefont{and}
  \bibinfo{author}{\bibfnamefont{R.}~\bibnamefont{Fr{\'e}sard}},
  \bibinfo{journal}{Physical Review B} \textbf{\bibinfo{volume}{73}},
  \bibinfo{pages}{33403} (\bibinfo{year}{2006}).

\bibitem[{\citenamefont{Koshibae et~al.}(2000)\citenamefont{Koshibae, Tsutsui,
  and Maekawa}}]{Koshibae:2000vj}
\bibinfo{author}{\bibfnamefont{W.}~\bibnamefont{Koshibae}},
  \bibinfo{author}{\bibfnamefont{K.}~\bibnamefont{Tsutsui}}, \bibnamefont{and}
  \bibinfo{author}{\bibfnamefont{S.}~\bibnamefont{Maekawa}},
  \bibinfo{journal}{Physical Review B} \textbf{\bibinfo{volume}{62}},
  \bibinfo{pages}{6869} (\bibinfo{year}{2000}).

\bibitem[{\citenamefont{Uchida et~al.}(2011)\citenamefont{Uchida, Oishi,
  Matsuo, Koshibae, Onose, Mori, Fujioka, Miyasaka, Maekawa, and
  Tokura}}]{Uchida:2011fv}
\bibinfo{author}{\bibfnamefont{M.}~\bibnamefont{Uchida}},
  \bibinfo{author}{\bibfnamefont{K.}~\bibnamefont{Oishi}},
  \bibinfo{author}{\bibfnamefont{M.}~\bibnamefont{Matsuo}},
  \bibinfo{author}{\bibfnamefont{W.}~\bibnamefont{Koshibae}},
  \bibinfo{author}{\bibfnamefont{Y.}~\bibnamefont{Onose}},
  \bibinfo{author}{\bibfnamefont{M.}~\bibnamefont{Mori}},
  \bibinfo{author}{\bibfnamefont{J.}~\bibnamefont{Fujioka}},
  \bibinfo{author}{\bibfnamefont{S.}~\bibnamefont{Miyasaka}},
  \bibinfo{author}{\bibfnamefont{S.}~\bibnamefont{Maekawa}}, \bibnamefont{and}
  \bibinfo{author}{\bibfnamefont{Y.}~\bibnamefont{Tokura}},
  \bibinfo{journal}{Physical Review B} \textbf{\bibinfo{volume}{83}},
  \bibinfo{pages}{165127} (\bibinfo{year}{2011}).

 \bibitem[{\citenamefont{Mukerjee and Moore}(2007)}]{Mukerjee:2007tg}
\bibinfo{author}{\bibfnamefont{S.}~\bibnamefont{Mukerjee}} \bibnamefont{and}
  \bibinfo{author}{\bibfnamefont{J.}~\bibnamefont{Moore}},
  \bibinfo{journal}{Applied Physics Letters} \textbf{\bibinfo{volume}{90}},
  \bibinfo{pages}{112107} (\bibinfo{year}{2007}). 

\bibitem[{\citenamefont{Morita	 et~al.}(2004)}]{Morita:2004ty}
\bibinfo{author}{\bibfnamefont{Y.}~\bibnamefont{Morita}},
  \bibinfo{author}{\bibfnamefont{J.}~\bibnamefont{Poulsen}},
  \bibinfo{author}{\bibfnamefont{K.}~\bibnamefont{Sakai}},
  \bibinfo{author}{\bibfnamefont{T.}~\bibnamefont{Motohashi}},
  \bibinfo{author}{\bibfnamefont{T.}~\bibnamefont{Fujii}},
  \bibinfo{author}{\bibfnamefont{I.}~\bibnamefont{Terasaki}},
  \bibinfo{author}{\bibfnamefont{H.}~\bibnamefont{Yamauchi}},\bibnamefont{and}
  \bibinfo{author}{\bibfnamefont{M.}~\bibnamefont{Karppinen}},
  \bibinfo{journal}{Journal of Solid State Chemistry} \textbf{\bibinfo{volume}{177}},
  \bibinfo{pages}{3149} (\bibinfo{year}{2004}).
  
\end{thebibliography}
\end{document}